\documentclass[manuscript,screen, nonacm]{acmart}
\usepackage{xcolor} 
\usepackage{colortbl} 
\usepackage{multirow} 
\usepackage{enumitem}
\usepackage[framemethod=tikz]{mdframed}

\newlength{\blocksize}
\newcolumntype{P}[1]{>{\centering\arraybackslash}p{#1}}

\AtBeginDocument{%
  \providecommand\BibTeX{{%
    \normalfont B\kern-0.5em{\scshape i\kern-0.25em b}\kern-0.8em\TeX}}}

\copyrightyear{2024}
\acmYear{2024}
\setcopyright{rightsretained}
\acmConference[CHI '24]{Proceedings of the CHI Conference on Human Factors in Computing Systems}{May 11--16, 2024}{Honolulu, HI, USA} \acmBooktitle{Proceedings of the CHI Conference on Human Factors in Computing Systems (CHI '24), May 11--16, 2024, Honolulu, HI, USA} \acmDOI{10.1145/3613904.3642118}
\acmISBN{979-8-4007-0330-0/24/05}

\begin{document}

\title{Simulating Wearable Urban Augmented Reality Experiences in VR: Lessons Learnt from Designing Two Future Urban Interfaces}


\author{Tram Thi Minh Tran}
\email{tram.tran@sydney.edu.au}
\orcid{0000-0002-4958-2465}
\affiliation{Design Lab, Sydney School of Architecture, Design and Planning,
  \institution{The University of Sydney}
  \city{Sydney}
  \state{NSW}
  \country{Australia}
}

\author{Callum Parker}
\email{callum.parker@sydney.edu.au}
\orcid{0000-0002-2173-9213}
\affiliation{Design Lab, Sydney School of Architecture, Design and Planning,
  \institution{The University of Sydney}
  \city{Sydney}
  \state{NSW}
  \country{Australia}
}

\author{Marius Hoggenmüller}
\email{marius.hoggenmuller@sydney.edu.au}
\orcid{0000-0002-8893-5729}
\affiliation{Design Lab, Sydney School of Architecture, Design and Planning,
  \institution{The University of Sydney}
  \city{Sydney}
  \state{NSW}
  \country{Australia}
}

\author{Luke Hespanhol}
\email{luke.hespanhol@sydney.edu.au}
\orcid{0000-0003-0839-481X}
\affiliation{Design Lab, Sydney School of Architecture, Design and Planning,
  \institution{The University of Sydney}
  \city{Sydney}
  \state{NSW}
  \country{Australia}
}

\author{Martin Tomitsch}
\email{Martin.Tomitsch@uts.edu.au}
\orcid{0000-0003-1998-2975}
\affiliation{Design Lab, Sydney School of Architecture, Design and Planning,
  \institution{The University of Sydney}
  \city{Sydney}
  \state{NSW}
  \country{Australia}
}

\renewcommand{\shortauthors}{Tran et al.}

\begin{abstract}
Augmented reality (AR) has the potential to fundamentally change how people engage with increasingly interactive urban environments. However, many challenges exist in designing and evaluating these new urban AR experiences, such as technical constraints and safety concerns associated with outdoor AR. We contribute to this domain by assessing the use of virtual reality (VR) for simulating wearable urban AR experiences, allowing participants to interact with future AR interfaces in a realistic, safe and controlled setting. This paper describes two wearable urban AR applications (pedestrian navigation and autonomous mobility) simulated in VR. Based on a thematic analysis of interview data collected across the two studies, we found that the VR simulation successfully elicited feedback on the functional benefits of AR concepts and the potential impact of urban contextual factors, such as safety concerns, attentional capacity, and social considerations. At the same time, we highlighted the limitations of this approach in terms of assessing the AR interface's visual quality and providing exhaustive contextual information. The paper concludes with recommendations for simulating wearable urban AR experiences in VR.
\end{abstract}




\maketitle

\section{Introduction}

Augmented reality (AR) is defined as a technology that blends computer-generated graphics with the real world in three dimensions (3D), enabling real-time interaction with virtual content~\cite{azuma1997survey}. Over the past two decades, AR research and development have gained significant momentum. AR experiences are now accessible to a broad audience due to the ubiquity of smartphones having the essential hardware requirement for AR and the availability of various development frameworks that simplify the creation of AR applications. The technology has also proven to be compelling to the public. In 2016, the location-based AR game Pokémon Go became a worldwide phenomenon~\cite{paavilainen2017pokemon}, demonstrating the potential of AR in general and urban AR more specifically. 

Although smartphones have facilitated the widespread adoption of AR, wearable devices, such as head-mounted displays (HMDs), are envisioned as involving a seamless integration of virtual imagery within an individual's field of view (FOV). The displays naturally augment a user's sensory perception of reality while allowing them to maintain an awareness of their real-world surroundings~\cite{azuma2019road}. Nonetheless, only a small number of AR studies have utilised HMDs in urban applications (e.g. navigation, tourism, and exploration)~\cite{dey2018systematic}. Technical challenges in making AR HMDs work outdoors, including ergonomic issues, insufficient display contrast, and unstable tracking systems~\cite{azuma1999challenge, billinghurst2021grand}, may have contributed to their under-exploration, compromising designers' ability to prototype and evaluate potential future applications. 

In this paper, we investigate virtual reality (VR) as a way to circumvent the challenges associated with prototyping AR experiences, thus bridging the gap between AR's current status quo and its potential application for urban interactions. VR has been shown to be an effective approach for simulating and evaluating different kinds of interfaces, from smart home products~\cite{Voit2019} and public displays~\cite{Maekela2020,yao2021evaluating} to yet-to-be-created technologies, such as fully autonomous vehicles (AVs)~\cite{shah2018airsim, tran2021review} and windscreen heads-up displays (HUDs)~\cite{colley2021effects, kim2009simulated, jose2016comparative}. When it comes to wearable urban AR concepts, augmentations are typically intended to relate meaningfully to the actual physical environment (e.g. navigation instructions overlaid onto the physical road). In these urban application scenarios, a VR simulation can reduce the complexity of creating conformal AR visualisations that require the precise alignment of virtual content and `real' physical objects, bypassing the need for world-scale AR technology \cite{billinghurst2021grand}. Moreover, many facets of the urban AR user experience are subject to contextual factors, such as social circumstances, temporal demands, and spatial settings. By using VR, it is possible to recreate and manipulate these aspects and, to a considerable extent, evoke realistic emotional~\cite{riva2007affective, Voit2019} and behavioural responses~\cite{deb2017efficacy, Maekela2020} from users. 

These advantages have led to studies utilising VR to simulate wearable urban AR interactions, allowing participants to experience idealised AR interfaces \textcolor{black}{in an immersive VR environment (AR-within-VR)}; for example, those that assist police officers in traffic stop operations~\cite{grandi2021design} and those that facilitate the interaction between AVs and pedestrians~\cite{prattico2021comparing}. To date, however, studies have yet to investigate the effectiveness of VR simulations in prototyping and eliciting relevant feedback about future urban AR experiences. As a prototyping method, the value of VR simulations is linked to the way they enable designers to `traverse the design space' and gain meaningful insights into the envisioned design~\cite{lim2008anatomy}. This use of VR simulations as a medium to `understand, explore or communicate what it might be like to engage with the product, space or system'~\cite{buchenau2000experience} is distinct from the practice of replicating AR systems, where the simulator's efficacy is demonstrated through comparable findings to real-world systems~\cite{bowman2012evaluating,lee2009replication, lee2010role}.

We contribute to the emerging domain of wearable urban AR through an analysis of two case studies: a map-based pedestrian navigation application and an interface supporting the interactions between pedestrians and AVs in future cities. These case studies are our own work, allowing for a thorough understanding of the design, prototyping, and evaluation process. Our analysis was driven by the following research questions (RQs): 
\begin{enumerate}
    \item To what extent can the method of simulating wearable urban AR experiences in VR elicit insights into the functional benefits of AR concepts and the impact of urban contextual factors?
    \item What are the limitations of simulating wearable urban AR experiences in VR?
\end{enumerate}

The paper makes two contributions. First, it determines the extent to which VR simulation helps participants to experience wearable AR in context and provide feedback on wearable urban AR experiences. Second, it provides a set of recommendations for simulating wearable urban AR experiences in VR, assisting researchers and practitioners in optimising their VR simulation.


\section{Related Work}

\subsection{Urban AR Applications}

Cities are evolving into complex ecosystems capable of utilising technology and data to tackle the daily challenges that residents are facing, improving their quality of life and the community's long-term resilience~\cite{ecosoc2015unece}. Urban interfaces, such as urban AR applications, have the potential to bridge the gap between individuals and the technological backbone of cities~\cite{tomitsch2017making} and have been explored across various urban domains, such as navigation~\cite{narzt2006augmented}, tourism~\cite{jingen2021systematic}, civic participation~\cite{parker2015keeping} and autonomous mobility~\cite{riegler2020research}. Most of them have been developed for smartphones, taking advantage of their omnipresence, hardware capacities, and established interaction paradigm. One recent example is Google Maps Live View\footnote{\url{https://blog.google/products/maps/take-your-next-destination-google-maps/}, last accessed January 2023}, which utilises the smartphone's camera to enhance localisation accuracy and projects directions in the real world. However, considering the urban context, handheld AR applications present one significant disadvantage: if not carefully designed, they might disengage users and place them at an increased risk of road accidents~\cite{simmons2020plight}. For example, pedestrians tend to use the AR navigation interfaces while walking~\cite{dunser2012exploring}; thus it is critical to promote a safer use\footnote{\url{https://support.google.com/maps/answer/9332056}, last accessed January 2023}. Therefore, with the ability to seamlessly integrate information into the user's view, wearable AR devices could become an even more attractive choice for urban applications.

Wearable AR technology is still in its early stages of development and has yet to attain operational maturity. However, it is essential to explore how we might design for compelling wearable urban AR experiences and address the potential contextual challenges associated with urban settings. According to \citet{rauschnabel2016augmented}, functional benefits are one of the most important drivers of wearable AR acceptance. They are defined as `the degree to which a user believes that using smart glasses enhances his or her performance in everyday life'~\cite{rauschnabel2015will}. Meanwhile, \citet{azuma2019road} suggested that consumer AR applications are relevant only when the combination of real and virtual content is substantially benefiting the overall experience. Therefore, one of the biggest challenges in designing wearable urban AR experiences is to maximise the perceived usefulness of digital augmentations. Other challenges concern the urban context in which wearable AR experiences are situated. The seemingly dense and messy urban environment could potentially interfere with the experience; for example, the constant flow of people passing by might interrupt the augmentation~\cite{javornik2018directions}. Complex traffic conditions in towns and cities pose various risks to pedestrians, such as falls, stumbles, and traffic collisions~\cite{oecd2012pedestrian}. Therefore, failure of an application to maintain the user's situational awareness might result in physical injuries~\cite{aromaa2020awareness}. Additionally, using AR glasses in public settings entails social acceptance issues for both users and bystanders~\cite{azuma2019road}. People are conscious about their image when wearing such glasses~\cite{rauschnabel2016augmented} and are hesitant to publicly use voice control or mid-air hand gestures~\cite{hsieh2016designing}. 

Through a rigorous analysis of user feedback for two representative wearable urban AR applications, this research determines whether these aspects of concerns (i.e. functional benefits, the impact of urban contextual factors) could be evaluated using a simulated environment. 

\subsection{Simulating Wearable AR Experiences in VR}

Prototypes are means for designers to gain valuable knowledge about different aspects of the final design~\cite{lim2008anatomy}. Because AR is fundamentally spatial, it is essential to depict 3D structures and relationships among design elements already early in the design process. Thereby, different AR prototyping methods of varying fidelity aim to incorporate some level of immersion in their artefacts, allowing participants to experience AR, interestingly, through a VR HMD or a Cave Automatic Virtual Environment. For rapid design exploration, sketches on equirectangular grids~\cite{nebeling2019360proto}, 360-degree panoramas ~\cite{pfeiffer2018exprotovar} and videos~\cite{berning2013parnorama} are becoming increasingly popular as a relatively lightweight approach to creating immersive AR interfaces. A number of specialised immersive authoring tools enable AR concept designs inside a virtual environment~\cite{freitas2020systematic} (e.g. Sketchbox3D\footnote{\url{https://design.sketchbox3d.com/}, last accessed January 2023},  Microsoft Maquette\footnote{\url{https://docs.microsoft.com/en-us/windows/mixed-reality/design/maquette/}, last accessed January 2023}, and Tvori\footnote{\url{https://tvori.co/}, last accessed January 2023}). These tools aim to lower the technical barriers, making spatial prototyping more accessible to creators. In the case of 3D game engine platforms such as Unity\footnote{\url{https://unity.com/}, last accessed January 2023}, the tools demand increased technical expertise, posing considerable challenges for non-programmers. However, they allow for more sophisticated AR interfaces while offering context manipulation, paving the way for context-aware or pervasive AR research~\cite{grubert2016towards}. This method---using high-fidelity VR simulation to prototype and evaluate wearable urban AR experiences---is, therefore, the focus of our paper. We discuss its advantages in greater detail below: 

\textit{Simulation of AR Systems:} A high-fidelity VR system features a powerful processor enabling low-latency positional tracking and a discrete graphics processing unit supporting high-resolution image rendering. Using high-end VR hardware and a software framework that allows independent manipulation of immersion parameters (e.g. FOV, latency, and stereoscopy), researchers can simulate different AR systems (including those yet to be developed) and investigate the effects of the isolated parameters in controlled studies~\cite{bowman2012evaluating}. Beyond AR system replication, VR hardware allows prototyping of immersive cross-reality systems~\cite{gruenefeld2022vrception} and future AR interfaces with a larger augmentable area and perfect registration of in-situ `AR' graphics~\cite{grandi2021design}. When used in conjunction with external input devices, VR also enables the prototyping of 3D interaction concepts~\cite{alce2015prototyping}. Furthermore, a simulated AR prototype can be rapidly iterated for numerous rounds of evaluation prior to deployment~\cite{burova2020utilizing, bailie2016implementing, gruenefeld2022vrception}. 

\textit{Simulation of Contextual Factors:} VR simulation provides control over various environmental factors, such as weather conditions and lighting. For example, \citet{gabbard2006effects} were able to reproduce the lighting conditions at night, dawn, and dusk in an attempt to investigate the effect of outdoor illuminance values on an AR text identification task. More significantly, the simulation approach allows for assessing AR systems in a broad range of dynamic situations that would be markedly hazardous, costly, or impossible to produce in a real setting. For example, VR simulation has been used to mitigate the risks associated with industrial maintenance~\cite{burova2020utilizing}, firefighting~\cite{grandi2021design,bailie2016implementing}, law enforcement~\cite{grandi2021design}, and AV–pedestrian interactions~\cite{prattico2021comparing}. Regarding context-aware AR interfaces designed to respond to contextual changes (e.g. information widgets change as a user goes from their home office to the kitchen), the virtual surroundings and even the accuracy in predicting user intent could be effectively simulated in VR~\cite{lu2022exploring}.

These benefits of VR simulation are of substantial relevance to the development of wearable urban AR applications, particularly in their ability to replicate an urban setting and its numerous contextual factors. According to \citet{hassenzahl2006user}, a good user experience promotes positive emotions and is concerned with the interactive encounter's dynamic, temporal and situatedness-related aspects. Therefore, an appropriate context, although simulated, may help to assess user experiences more holistically. Until recently, however, there has been little knowledge on utilising VR simulation to evaluate wearable urban AR applications. Thus, an assessment of the method’s efficacy in obtaining insightful user feedback is lacking, and factors that should be considered when simulating wearable AR experiences in VR remain unclear. The research described in this paper aims to address these gaps and enhance the prospects of more wearable urban AR applications being prototyped and evaluated in context.


\section{Case Studies}

\begin{table}[h]
  \caption{Prototype characteristics and participants' demographics in each case study.}
  \label{tab:casesummary}
  \begin{tabular}{lcc}
    \toprule
    &\textbf{Pedestrian Navigation}&\textbf{AV Interaction}\\
    \midrule
    \textbf{Number of Conditions} & 3 & 4\\
    \textbf{VR Exposure per Condition} & 3-5 minutes & 1-1.5 minutes\\
    \textbf{Movement} & Joystick-based & \textcolor{black}{Real} walking\\
    \textbf{Interaction} & Controller & Hand gesture \\
    \textbf{AR Content} & Maps, turn arrow & Text prompt, crossing cues \\
    \midrule
    \textbf{Number of Participants (m/f)} & 18 (9/9) & 24 (9/15)\\
    \textbf{Previous VR Experience}\\
    Never &2 &8\\
    Less than 5 times &15 &13\\
    More than 5 times &1 &3\\
    \textbf{Study Location} & Australia & Vietnam\\
  \bottomrule
\end{tabular}
\end{table}

The following sections describe two case studies involving the design and evaluation of two wearable urban AR applications, both of which were developed using immersive VR as a means for prototyping. In the first case, we evaluated an AR pedestrian navigation application featuring different exocentric map displays. In the second case, we evaluated a wearable AR application assisting crossing pedestrians in autonomous traffic. Despite sharing the urban context, these applications were conceptualised, prototyped, and evaluated at different times and independently of each other. Both studies were led by the first author of this paper. Table \ref{tab:casesummary} shows prototype characteristics and participants' demographics in each case study. It is worth emphasising that these studies were designed to assess their respective AR concept prototypes, whose results were reported in prior publications~\cite{thi2020designing, tran2022designing}. In this paper, we analysed the qualitative data obtained from both studies together to demonstrate the efficacy of VR in simulating wearable AR experiences in a generalised manner.

\subsection{Pedestrian Navigation}

\subsubsection{Design Context} 
Urban wayfinding has always been a challenge due to the dense network of streets and buildings. In recent years, a substantial amount of work has examined how AR could improve guidance by superimposing directional cues on the real world~\cite{narzt2006augmented}. However, few studies have utilised AR HMDs despite hands-free devices potentially providing a more seamless experience. The narrow FOV of existing AR HMDs also restricts the amount of information that can be displayed and searched through~\cite{trepkowski2019effect}, contributing to the relatively sparse research on map-based navigation compared to turn-by-turn directions~\cite{lee2022user, zhao2020effectiveness}. It is expected that a wider FOV would enable improved freedom in displaying high-level spatial information to users. Thus, determining the influence of topographic map placement on user experience and navigation performance is of relevance. 

\begin{figure}[t]
\centering
  \includegraphics[width=1\linewidth]{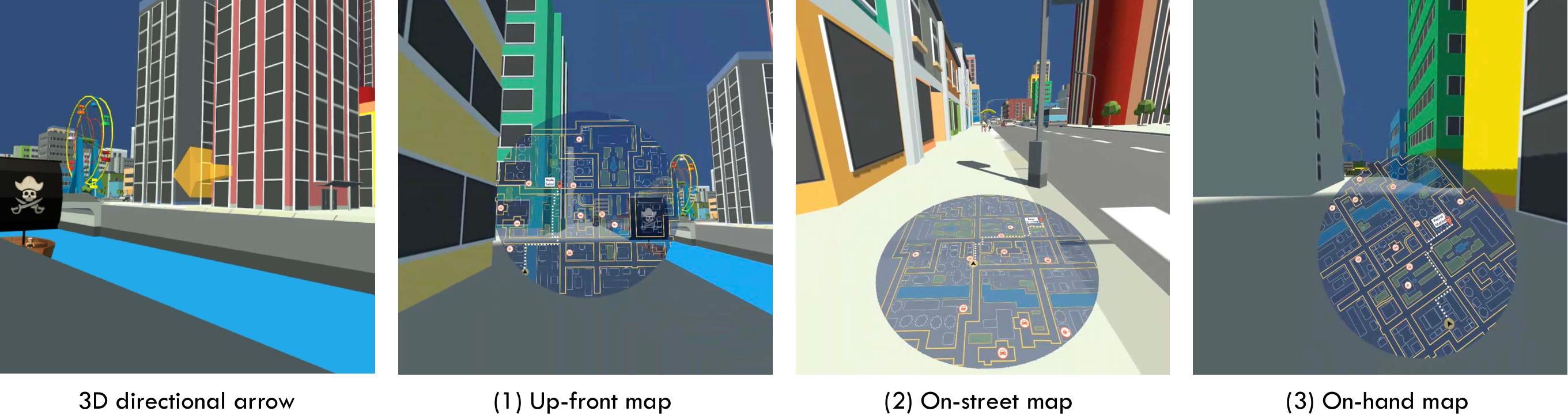}
  \caption{The AR application to support pedestrian navigation featured a directional arrow to display turn directions when the map was not in use (as shown in the far left image). For the AR map view, we investigated three different map positions: up-front map (1), on-street map (2) and on-hand map (3).}
  \label{fig:NavConditions}
\end{figure}

In this case study, we examined three distinct map positions, namely (1) at a distance in front of the users, (2) on the surface of the street, and (3) on the user's hand (Figure \ref{fig:NavConditions}). Except for the on-hand map, which was anchored to the right hand and visible only when the user brought it up, the other two maps followed the user's visual field once activated. These different map positions were inspired by previous works on AR HMD maps in the industry (e.g. Google Glass~\cite{joy_2013}) and academic research \cite{goldiez2007effects, lehikoinen2002walkmap}. Following the recommendation by \citet{goldiez2007effects}, we offered users the flexibility to access the maps \textit{on demand} to avoid obstructing views and gain insights into their map usage. The VR simulation allowed users to experience an idealised interface that was not bound by existing AR constraints (e.g. limited FOV, insufficient brightness and contrast, unstable large-scale tracking, and lack of GPS sensors) in a setting similar to the real world. Concerns such as pedestrian safety and having control over the experimental setting were also major factors in the decision to use VR.

\subsubsection{Prototype Development}

The prototype was created using the Unity game engine and experienced with an Oculus Quest 1 VR headset. The selected VR system offered inside-out tracking, a display resolution of 1440 x 1600 pixels per eye and built-in audio~\cite{quest1}. We designed a virtual city (Figure \ref{fig:NavSetup}) using pre-made 3D models from the Unity Asset Store\footnote{\url{https://assetstore.unity.com/}, last accessed January 2023}. In order to create a lively urban atmosphere for the city, we simulated road traffic and various activities along navigation routes (e.g. people waiting at bus stops or police pursuing criminals). Additionally, ambient street sounds were added to the virtual scenes and configured to support spatial sound perception.

To create the map interface, we captured the orthographic view of the virtual city in Unity. The image was then imported into Adobe Photoshop for further editing. Specifically, we reduced the amount of information encoded in the map to avoid unnecessary details and made it semi-transparent. A white dotted line indicated the recommended route, and a black arrow denoted the user's current position, with a pointer indicating their facing direction. In addition to the map, the navigation application incorporated egocentric turn-by-turn guidance represented by a 3D arrow. While navigating, participants could switch between the map and arrow views by pressing a button.

Participants traversed the virtual environment using the controllers' joysticks. This locomotion technique was implemented to facilitate long-distance travel without requiring specialised hardware (e.g. an omnidirectional treadmill~\cite{souman2011cyberwalk, jayaraman2018trust}). Compared to other techniques such as teleportation, joysticks are easier to use \cite{boletsis2019vr} and allow participants sufficient time to observe the surroundings. However, they generally induce more motion sickness \cite{di2021locomotion}, necessitating the use of a constant and slow-moving speed to minimise nausea. Participants were also instructed to remain seated throughout the VR experience as a safety precaution. Due to the lack of ecological validity of artificial locomotion, the study examined only the user's augmented visual immersion in the environment and less about their bodily immersion.


\subsubsection{Evaluation Study}
\hfill\break
\textit{Experimental Design.} We conducted our user study in a within-subjects design with three conditions (three map placements) presented in a counterbalanced order. Participants were asked to navigate to a pre-determined destination (e.g. a petrol station) and to locate a specific location (e.g. the nearest bus stop).

\textit{Participants and Location.} We recruited 18 participants from the university campus and the neighbouring area using the university's mailing lists, our social media networks, and word of mouth. Nine participants self-identified as male and nine as female (aged 18-34). All participants spoke English fluently and had normal or corrected-to-normal eyesight. Prior to the study, the majority of participants stated that they had little to no experience with VR technology. Only one participant had extensive experience with VR as a result of her enrolment in a VR development course. Participants were all familiar with digital navigation applications (e.g. Google Maps). The study was carried out in our lab in Australia. 

\textit{Study Procedure.} After briefing participants about the navigation study, we asked them to read and sign a consent form. We then assisted participants in wearing the headset and adjusting the head strap to their comfort. This step was followed by an interactive tutorial designed to mitigate the novelty effect of VR~\cite{miguel2022developing}, where participants learnt to operate the controllers and traverse the virtual city. Following each condition, participants were asked to complete a set of standardised questionnaires and a general feedback form. Upon completion of the study, participants were invited to partake in a semi-structured interview. The experiment lasted for a maximum of 75 minutes, including the time required for participants to rest between conditions. Each interface was experienced for approximately 3 to 5 minutes (similar to other map navigation studies, e.g.~\cite{goldiez2007effects}).

\textit{Data Collection.} Along with questionnaire data and HMD-logged data such as task completion times, we collected qualitative data using the post-trial general feedback form (probing favourable aspects and areas for improvement) and post-study semi-structured interviews. The interview aimed to learn about {(1) participants' overall experience, (2) their preferred conditions and (3) their opinions about different aspects of the prototype, such as interaction and information display. Participants were also asked about their VR experience and whether they suffered from motion sickness. 

\begin{figure}[h]
  \centering
  \includegraphics[width=0.7\linewidth]{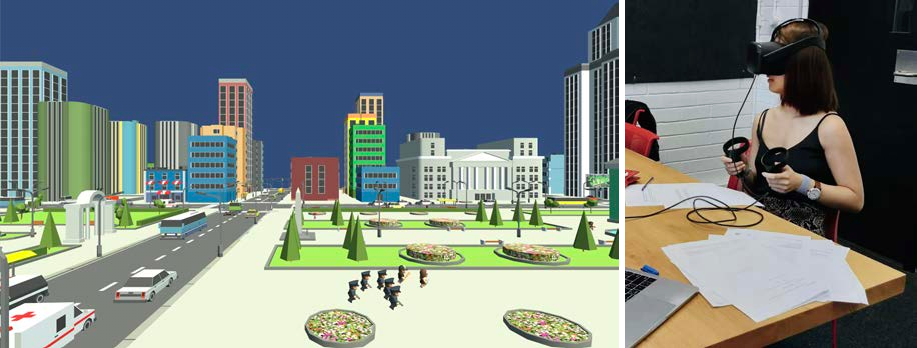}
  \caption{The simulated environment of the navigation study (left) and a participant using controllers for movement and interactions (right).}
  \label{fig:NavSetup}
\end{figure}

\subsection{Interaction with Autonomous Vehicles}

\subsubsection{Design Context.} 
Future AVs may be equipped with external human-machine interfaces (eHMIs) to compensate for the absence of driver cues. It has been shown that the presence of the eHMIs makes it easier for pedestrians to grasp an AV's intent and prevents ambiguous situations~\cite{rouchitsas2019external}. Most of these external display technologies are integrated into or attached to the vehicle, utilising its outer surface (e.g. windshields) or nearby surroundings (e.g. laser projection). Other possible placements include road infrastructures and wearable devices~\cite{dey2020taming}. Faced with scalability challenges arising from multi-agent situations~\cite{colley2020unveiling}, researchers have recently shown an increased interest in wearable AR for its potential in assisting pedestrians with highly personalised and contextual information~\cite{tabone2021vulnerable}. To date, there have not been any studies examining AR concepts in heavy traffic, and evaluations of AR have primarily focused on comparing holographic visualisations~\cite{hesenius2018don, prattico2021comparing, tabone2022augmented} rather than on whether pedestrians would prefer a wearable AR solution. 

To address this gap, this case study investigated the extent to which users prefer to use AR glasses for sending crossing requests to approaching AVs compared to a traditional pedestrian crossing button. This AR design concept represents an infrastructure-free method that may become feasible in the advent of the Virtual Traffic Lights system (as detailed in the following patent~\cite{tonguz2021system}). Additionally, we examined the impact of three communication approaches---a visual cue being placed on (1) each vehicle (distributed) or (2) the street (aggregated) or (3) both---on pedestrians' perceived cognitive load and trust. Utilising VR, we were able to construct a complex traffic scenario with many AVs travelling down a two-way street. In the real world, such a scenario would have been impossible due to the possibility of causing physical harm to both participants and researchers. 


\begin{figure}[t]
\centering
  \includegraphics[width=1\linewidth]{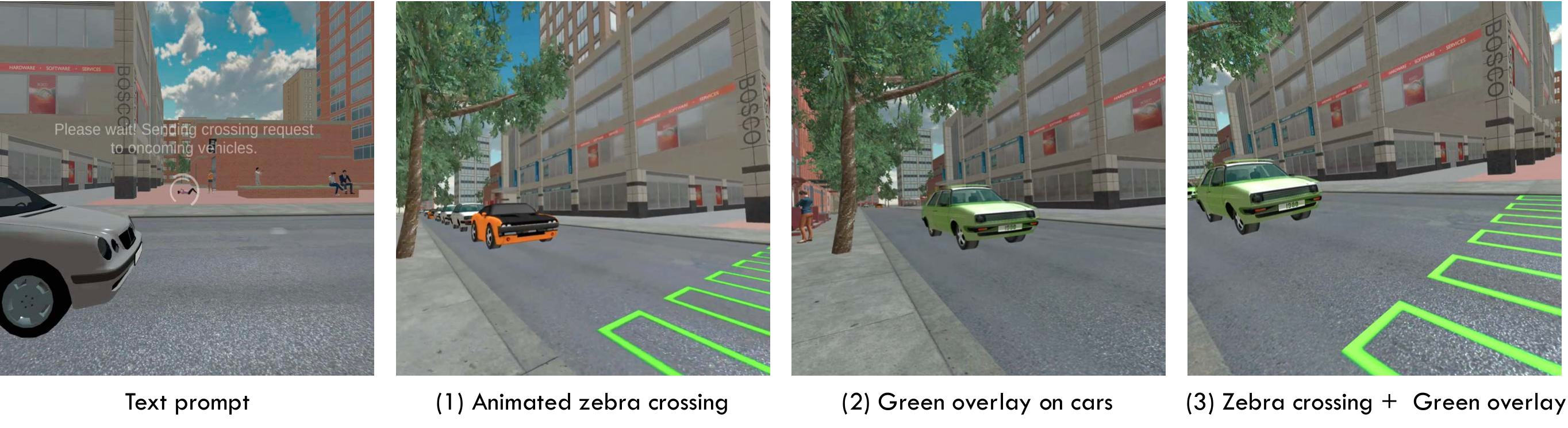}
  \caption{A text prompt indicating that the crossing request was successfully received by the AR application (far left). To indicate to pedestrians when it is safe to cross, we compared three different AR visual cues: animated zebra crossing (1), green overlay on cars (2) and a combination of both (3).}
  \label{fig:Avconditions}
\end{figure}

\subsubsection{Prototype Development.}

We employed Oculus Quest 2, a standalone VR system, to allow participants to freely move around unconstrained by cables and cords. Its hand-tracking feature also enabled us to prototype hand gestures without the need for additional sensors. The virtual environment was developed in Unity. We used off-the-shelf 3D models from the Unity Asset Store to create an urban road with two lanes and two-way traffic (Figures \ref{fig:Avconditions} and \ref{fig:AVSetup}). Traffic was a random mix of three vehicle types: a black/orange sports car, a silver sedan and a white hatchback, all of which travelled at approximately 30 km/h. The scene also included several lifelike 3D characters obtained from Mixamo\footnote{\url{https://www.mixamo.com/}, last accessed January 2023}, engaging in various activities such as exercises and talking. These virtual persons were carefully situated at a distance from participants to avoid causing distraction~\cite{hoggenmuller2021context}. To further enhance participants' sense of presence in VR, we included an ambient urban soundscape with bird chirps and street noise. Additionally, we used sounds designed by Audi\footnote{\url{https://www.e-tron-gt.audi/en/e-sound-13626}, last accessed January 2023} for electric vehicles and adjusted the Unity Spatial Blend parameter to simulate 3D sounds. 

In the simulated environment, participants could see and use their virtual hands to press the pedestrian button or activate the AR application by tapping on the VR headset (Figure \ref{fig:AVSetup}). The latter interaction was made possible by enclosing the headset in an invisible collision zone that detects any contact with the user's fingertips. The AR application featured three different holographic visualisations: a zebra crossing, a green overlay on each vehicle and a text prompt that informed users of the application's action and instructed them to wait. We used a semi-transparent white colour for both the text and loading animation, aiming to imitate a futuristic light-based interface. The zebra crossing's appearance and animation were modelled after Mercedes-Benz F 015's projected crosswalk\footnote{\url{https://youtu.be/fvtlobbMENo}, last accessed January 2023}. The car overlay was a duplicated version of the car frame, with the mesh's material changed to green neon paint and the opacity set at 100. The overlay was also enlarged to appear slightly detached from the car. Along with visual elements, we included audio to accentuate specific events. In particular, successful activation of the application was signalled by a bell sound, slow chirps were used to indicate waiting time, and a rapid tick-tock sound meant crossing time. These audio cues were obtained from the PB/5 pedestrian button because the spoken texts of the original button were deemed unsuitable as \textcolor{black}{user interface (UI)} sounds.

\subsubsection{Evaluation Study}
\hfill\break
\textit{Experimental Design.} The study was designed as a within-subjects experiment with four conditions: a baseline (the pedestrian button) and three variations of the AR application (Figure \ref{fig:Avconditions}). Each experimental condition began with the participants standing on the sidewalk. Their task was to use the provided system and physically cross the road. 

\textit{Participants and Location.} A total of 24 participants (aged 18-34) took part in the study, of which nine self-identified as male and fifteen as female. Participants consisted of professionals and university students interested in the topic. All participants were recruited through social media networks and word of mouth. To participate, they were required to have (corrected to) normal eyesight, normal mobility, good command of English and to have been living in the city where the study took place for at least one year. Most participants had little to no previous experience with VR or AR. The study took place at a shared workspace for technology startups in Vietnam.

\textit{Study Procedure.} Following a briefing and the signing of a consent form, we invited participants to stand in the starting position and put on the headset. An instructional session was designed for participants to familiarise themselves with the immersive virtual environment, learn how to use hand-based interactions and practice crossing the street. The physical movement area was approximately 3 by 8 metres. Before each experimental condition, participants were informed about the technology used (either AR glasses or a pedestrian button). Following each condition, participants were instructed to remove the headset and to answer a series of standardised questionnaires. After having completed all the conditions, participants were asked about their experiences in a semi-structured interview. The study took about 60 minutes to complete. Considering previous VR studies investigating pedestrian behaviour~\cite{schneider2020virtually}, we decided on a short task duration (approximately 1 to 1.5 minutes) to ensure that participants are sufficiently immersed in the scenario without becoming bored or fatigued from repeated crossings.

\textit{Data Collection.} In addition to questionnaire data, we obtained qualitative data relevant to the research questions addressed in this paper using post-study semi-structured interviews. Participants were inquired about (1) the overall experience, (2) their preferred conditions and (3) their opinions about aspects such as interaction, information display and simulated environment. 

\begin{figure}[h]
  \centering
  \includegraphics[width=0.7\linewidth]{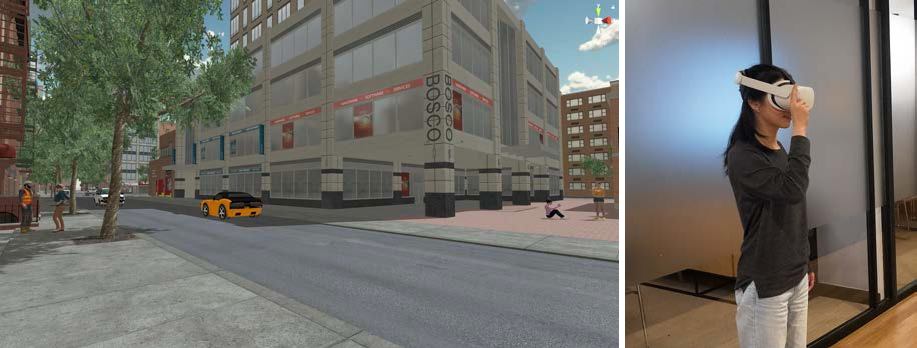}
  \caption{The simulated environment used in the AV study (left) and a participant tap\textcolor{black}{p}ing on the headset to send a crossing request (right).}
  \label{fig:AVSetup}
\end{figure}

\section{Data Analysis}

The qualitative and reflective analysis presented in this paper seeks to understand how participants experienced and valued different aspects of the AR prototypes. Furthermore, it examines participants' sense of presence in the virtual environment and how it influenced their emotional and behavioural responses. 

Post-study interviews from the two case studies were transcribed by speech-to-text software and then reviewed and edited by the interviewer. We applied an inductive thematic analysis method~\cite{braun2006using} to analyse the data, using digital sticky notes on Miro. The analysis was performed by two coders with different levels of engagement throughout the studies. Whereas the first coder designed and performed both studies, the second coder was not involved in their implementation. The process involved the first coder reading through all data and selecting a representative subset (4/18 interviews for the navigation study, 5/24 interviews for the AV study). Both coders worked on the same subset of interviews independently, followed by a discussion to review differences and to finalise the code book. The first coder then applied the agreed coding frame to all interviews. Any new data points discovered during this process were discussed with the second coder. All identified themes and sub-themes were reviewed by the research team and formed the basis of the Results section. 


\section{Results}

\subsection{Participants' Feedback on AR Prototypes} 

In this section, participants from the navigation study are denoted as N (e.g. N18) and from the AV study as A (e.g. A18). The number of participants from each study is indicated by \emph{navi} and \emph{av}, respectively (e.g. navi=6, av=3).

Figure \ref{fig:v1} summarises different aspects of the AR prototypes about which the participants provided feedback. It also illustrates the differences between the two case studies regarding the type of user feedback and quantities. 

\subsubsection{AR Glasses} 

As the majority of our participants had no prior experience interacting with AR glasses, their opinions regarding the technology and its potential adoption were based solely on their direct experiences with the prototypes. In our analysis, participants commented on different factors influencing the adoption of AR glasses, such as their eyewear nature (navi=3, av=4), the unfamiliarity of AR technology (av=4) and potentially high cost (av=4). Because AR glasses are individually owned---in contrast to a public solution, such as the pedestrian button used in one of our studies---a few participants mentioned the inconvenience of forgetting their glasses at home (av=2), and another two pointed out that they would need to bear the responsibility of taking care of the device (av=2). Of note, three participants found the device to not be absolutely necessary for the street crossing task (av=3) and the AR glasses were only \textit{`nice to have'} (A6). Alternative technologies, such as smartphones, were mentioned by several participants (navi=1, av=3).

Four participants made reference to the potentially larger ecosystem of applications available on such AR glasses (av=4): as a multi-purpose device, the AR glasses appealed to some (av=3), but at the same time, there existed concerns about the potential advertisements or disconnection from the physical world. For example, A9 commented, \textit{`You can also read news and watch TV, but I am afraid that we lose connection in the real world'}. Sharing a different perspective, A1 believed that AR glasses might help users retain their attention compared to using mobile devices, \textit{`I can't browse social media feeds, since it's not that great to do that on the AR glasses'}.

\subsubsection{Functionality} 

Prototyping plays a critical role in validating and informing design concepts in the early phase of product design~\cite{lim2008anatomy}. Our analysis revealed that a number of participants articulated precisely which aspects of the design concepts they found most valuable (navi=9, av=6). In the navigation study, participants commended the blending of spatial directions into the natural environment, mentioning that it fixed the issue of not knowing which direction they were facing (navi=4). Further, they commented that the design concept would be particularly useful for navigating through unfamiliar places or dense areas with complex walkways (navi=5). In the AV study, participants emphasised the convenience of sending crossing requests to vehicles and being able to safely cross the street at any location (av=6). A6, for example, mentioned her negative experiences with impatient drivers, \textit{`I'm very concerned about my safety when crossing the street, but now I trust that when I choose the ``activate'' mode to send signals to all of the vehicles, they will all stop, and I will have time to cross the street'.} Nevertheless, the perceived usefulness of the application was less about graphical augmentations. Instead, participants appreciated the flexibility in choosing their crossing locations and having control over the interaction, both of which smartphones could, however, readily provide. 

More than half of the participants in the AV study (av=13) considered the complex real-world situations when providing feedback about the design concept. They commented on the feasibility of sending crossing requests in mixed traffic scenarios (av=5) in which \textit{`the traffic is still filled with a lot of manual [vehicles], and motorbikes'} (A2). Two participants also mentioned local compliance with traffic laws as a factor influencing their trust in the implementation (av=2). A1, for example, stated, \textit{`In the EU, I will trust it, but here in Vietnam, I doubt it'.} Six participants questioned whether misuse or an increase in the number of crossing requests would negatively impact traffic efficiency (av=6). Lastly, because the concept involves communication between AR glasses and vehicles, several participants were suspicious about the handling of their personal data (av=4).

\subsubsection{User Experience}
We discovered various evaluative statements concerning the pragmatic qualities of the prototypes in both case studies. In regard to positive aspects, participants described the AR applications as \textit{`useful'}, \textit{`easy to use'}, \textit{`convenient'}, \textit{`well integrated'} and \textit{`intuitive'} (navi=12, av=4). Meanwhile, participants' sense of safety appeared to be a deciding factor in their preference towards a specific version of the application. Many participants in the navigation study were particularly cautious about how augmented navigational instructions might interfere with concurrent activities, such as sight-seeing and paying attention to the streets (navi=15). N7, for example, complained about the opaque up-front map, \textit{`I couldn't see where I was going while using the map [...] I had to stop or keep walking in an open area'.} As a result, they preferred a safer design solution for mobile use; one could better maintain their situational awareness, \textit{`While I was following the arrow, I was actually observing people around, [...] I saw a car making a U-turn. There were some policemen running after a woman' (N1)}. Whereas safety was perceived as one of the design requirements for the navigation application, participants considered safety as the most critical aspect when interacting with AVs. They evaluated different conditions based on their subjective feelings when crossing the street regarding whether they would feel \textit{`safe'}, \textit{`confident'}, \textit{`uneasy'}, \textit{`insecure'}, \textit{`rushed'} or \textit{`worried'} (av=11). Hedonic aspects occurred rarely in the qualitative data, even though the AR applications were sometimes described as \textit{`cool'}, \textit{`exciting'}, \textit{`awesome'} and \textit{`impressive'} (navi=3, av=3). 

\subsubsection{Information}

A large number of participants (navi=7, av=23) provided feedback on the usefulness of different message types. Several statements, interestingly, revealed differences between the designer's intention and the user's interpretation. For example, the zebra crossing was intended to be a visual cue indicating when the user should cross. However, we found that participants also relied on the crosswalk to recognise the safe crossing zone and the boundary where the vehicles would stop (av=5). In the conditions without the crosswalk, they felt less confident. One participant, A16, even decided not to cross, \textit{`I just do not know where [the cars] will stop, inside or outside the area. They may stop very close to me'}. Regarding the number of visual cues, most participants in the AV study felt assured receiving multiple indications to cross given the dangerous nature of road traffic (av=12). For A18, \textit{`it's like double the security'}. Meanwhile, two participants were concerned about getting distracted by multiple visual (A6) and audible (A21) cues. 

In the navigation study, the amount of information presented on the map was expected to be \textit{`minimal'} (N7) yet sufficient for the task at hand (navi=8). Further, our analysis recorded a large number of suggestions for additional information (navi=15, av=11), uncovering what participants thought was missing from the applications. Participants in the navigation study desired cardinal directions (navi=5), estimated time or distance to arrival (navi=7), information about nearby locations (navi=4), upcoming turns (navi=5) and warnings (navi=3). In the AV study, the application was requested to display waiting time (av=3), crossing time (av=7), the number of vehicles that had received the crossing request (av=5), and notification of task completion (av=1). It is worth noting that among the feedback, a variety of proposals for new features were offered, e.g. oriented search, a navigation approach in which users rely on distal visual cues to orient themselves and work out the direction to the destination~\cite{golledge1999wayfinding}. N1 suggested, \textit{`This is AR, right? You could add something to the location of the destination. If I am going to the Four Seasons Hotel, there could be something in my sight that I can see from a far distance throughout the process'}.

\subsubsection{Visualisation} 

There was a wide variety of user feedback pertaining to the visualisation of AR content. In terms of recognisability, a large number of participants in the AV study failed to notice the car overlay (av=10). Meanwhile, we did not observe a similar issue with the zebra crossing, even though it was also designed to be part of the real world. According to the user feedback, the problem could be attributed to two factors. First, the participants did not notice when the overlay appeared due to the sheer number of moving vehicles. For example, A21 stated, \textit{`There were so many cars on the road, [...] I had to turn left and right, and my attention was divided'}. Second, they could not distinguish the AR overlay from the car itself, thinking \textit{`[it was] just a green car'} (A20). Of note, one participant stated that AR overlays affected his impression of real-world objects, \textit{`the vehicles […] were more futuristic, probably because of the overlay'} (A1). Regarding clarity or comprehensibility, participants expressed preferences for short and straightforward instructions (av=7) combined with relevant icons (av=3). In indicating when it is safe to cross, five participants stated that a text prompt might convey the message more clearly than graphical representations (av=5). 

With respect to the visual appearance of the AR content, the following characteristics were mentioned: aesthetics (navi=6), colour (navi=3, av=5), transparency (navi=7), size (navi=8), and animation (navi=1, av=1). It is worth noting how those visual aspects might divert users' attention (navi=2, av=1). As N7 expressed, \textit{`the vertical map is somehow transparent but still distracts me from the streets.'} Besides, moving or rotating overlays had to be used with caution because two participants reported their potential in causing sickness (navi=1, av=1). 


\begin{figure}
\centering
  \includegraphics[width=0.9\linewidth]{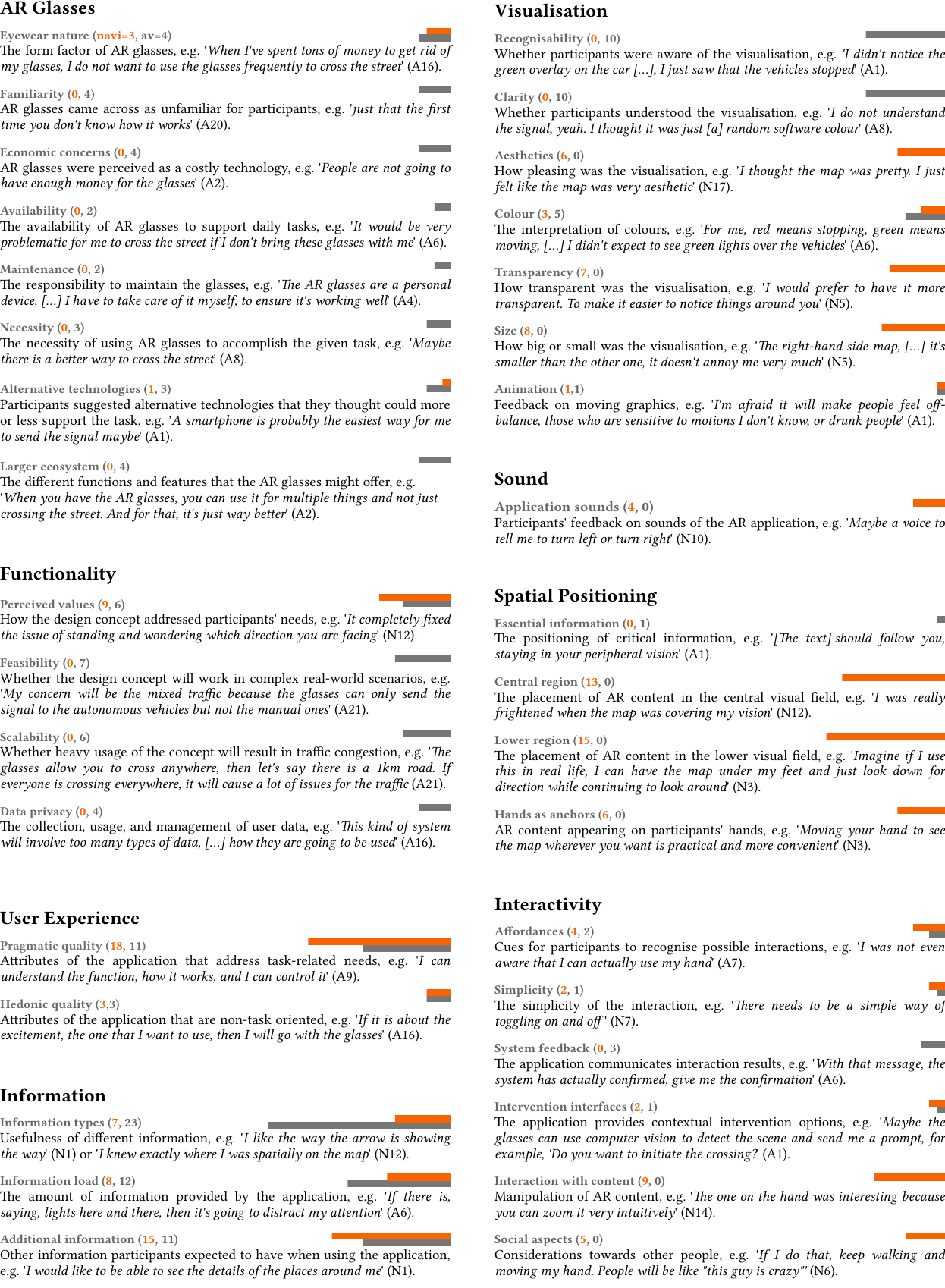}
  \caption{Different aspects of the AR prototypes about which the participants provided feedback. The orange bar represents the navigation study; the grey bar represents the AV study. The bar length represents the number of participants who provided feedback for a specific aspect for the navigation study and the AV study, respectively.}
  \label{fig:v1}
\end{figure}

\subsubsection{Sound}
In the navigation study, we found several suggestions to incorporate voice-guided navigation (navi=4). In the AV study, although different sounds accompanied user interactions, there was no feedback related to the auditory aspect. Interestingly, in the conditions in which crossing was supported through the AR glasses, participants seemed to take little notice of the sounds. On the other hand, when the same sounds were applied to interacting with the pedestrian button, five participants felt the pressure to cross the street as quickly as possible (av=5). For A1, the sounds felt like \textit{`an alarm clock'}. Meanwhile, for A16, the sound was similar to \textit{`a countdown'.}

\subsubsection{Spatial Positioning} 
AR content can be placed anywhere in the 3D environment, making it challenging for designers to choose an optimal position. A1 stated that essential information, such as crossing indication, should be rendered within the user's field of vision rather than situated in the world space. He reasoned that \textit{`both the zebra crossing and the vehicle's colour could only be seen when you look at the street. When you look in a different direction, you won't see them anymore'.} Meanwhile, placing content that might be used en route for an extended period (e.g. navigational interface) in the central view was reported to cover real-world objects (navi=10) and divert users' attention (navi=4). This feedback concerned not only the map (navi=8) but also the relatively small directional arrow (navi=4). N9, for example, expressed her frustration that \textit{`the arrow in the middle [of the view] made [her] feel annoyed'.} When content was placed in the lower region (i.e. on the street), despite being less obstructive (navi=10), it led to issues such as divided attention (navi=5), neck strain (navi=3), and suboptimal text legibility (navi=2). Additionally, three participants worried about not being able to spot obstacles on the ground (navi=3). Anchoring the map to the user's hand was perceived as natural (navi=6). N10 was reminded of when \textit{`[he] used Google Maps'} on his smartphone while walking, and felt that \textit{`the experience [was] quite real'.} Meanwhile, N14 found it safe when the on-hand map did not hinder his vision, and he only referred to it when required, explaining that \textit{`you're not working like this with your hand. It's very unnatural. So you need to put your hand down'.} Four participants, however, mentioned that the on-hand map position was strange and \textit{`awkward'} (N14) (navi=4).

\subsubsection{Interactivity} 

Several participants were unsatisfied about the absence of affordances for interaction, stating that they had received no hints regarding possible engagement with the AR glasses (navi=4, av=2). Three participants reflected on the simplicity of interactions (navi=2, av=1). A1, for example, mentioned the ease of \textit{`tapping on the glasses'} as opposed to \textit{`going through many menus'.} At the same time, the same participant was concerned about the possibility of accidental interaction. System feedback to interaction requests was perceived as necessary by several participants (av=3). For example, A1 assumed, \textit{`If I don't see the message, I suppose I will press it a lot of times'.} 

A few participants expected the AR glasses to sense the environment and suggest relevant interventions~\cite{schmidt2017intervention} (navi=2, av=1). A4, for example, wished to be notified if crossing the street at the wrong spot and taking the wrong route. In the navigation study, the analysis revealed a large amount of feedback relating to interacting with the holographic map (navi=9). Participants appreciated having physical control over the positioning of content (navi=4). N15 stated, \textit{`I like that you can control the map; it is like you are literally holding it'.} Five participants mentioned the opportunity to view the content in multiple ways. In particular, they expected to extract relevant information by zooming (navi=4) and filtering (navi=2). Social aspects of the interaction were mentioned by several participants (navi=5). For N6, moving his hands while walking would be considered unusual, \textit{`People will be like, "this guy is crazy"'}. Meanwhile, N7 was concerned more about how she might accidentally hurt nearby people, \textit{`obviously you couldn't use [the on-hand map] when it’s close to people because you might hit them with your hand'}.

\subsection{Participants' Behaviour in Immersive Virtual Environments} 

Joystick-based navigation resulted in more motion sickness feedback than \textcolor{black}{real} walking (navi=12, av=3); however, the discomfort was reported to be mild and decreased over time. The VR simulation used for the AV study presented an upgrade in image resolution (Oculus Quest 2 versus Oculus Quest 1) and visual fidelity (high-poly versus low-poly models). Interestingly, it received less positive feedback regarding perceived realism than the navigation study (navi=8, av=5). Ambient sounds of people talking (navi=6), familiar scenes (av=2), and various social activities in the background (navi=2, av=2) had a noticeable impact on participants \textit{`feeling real'}. For example, N5 felt as though she was \textit{`walking in a real city'} due to the sounds of people talking, while N7 expressed a fondness for the \textit{`similarities to a real-life city'}. We further found that the perceived realism of the experience triggered strong emotional and behavioural responses among the participants; these reactions were notably evident in the encounters with virtual traffic. The participants exercised caution around the cars (navi=1, av=19) despite being aware of VR as a safe environment. For example, in the navigation study, N13 stated, \textit{`I know that those cars might not harm me because they are not real, but I have the instinct to run away, [...] when the car was near me I almost jumped.'}. Meanwhile, participants in the AV study demonstrated typical pedestrian behaviour, such as hesitating to cross, waiting for vehicles to come to a complete stop, or rushing across the street. Nevertheless, the simulations could not completely recreate the various sensory inputs prevalent in the real world, as revealed by participants in both studies (navi=3, av=6). In this regard, A21 believed that there would be \textit{`more sensory inputs'} that could influence his crossing decisions in real life.


\section{Discussion}

In this section, we first review the user feedback obtained and discuss the extent to which VR simulation elicit insights that are valuable for the design of wearable urban AR experiences, thereby addressing RQ1. We organise this section around the challenges for designing wearable urban AR applications: (1) functional benefits and (2) impact of urban contextual factors, as detailed in the Related Work section. Next, we address RQ2 by discussing the limitations of VR simulation. Finally, we present a set of recommendations for simulating wearable urban AR experiences in VR. The discussion concludes with the acknowledgement of the study's limitations.

\subsection{VR Simulation Efficacy in Evaluating Wearable Urban AR Applications (RQ1)}

Undoubtedly, simulations were unable to reproduce the physical characteristics of the AR glasses (e.g. size, weight, and aesthetics); any feedback related to these characteristics was directed towards the VR headset. However, one can postulate that to a certain extent VR HMD could give participants the impression of wearing smartglasses, thus helping them to envision better how AR technology could become part of their daily lives. For instance, AR glasses were not appealing to some participants in the AV study, but the reason was less about their physical form factors, as previously discovered~\cite{rauschnabel2016fashion}, and more about the fact that participants were obliged to wear glasses. This concern was especially prevalent among people who underwent medical procedures to correct vision problems (e.g. astigmatism, nearsightedness, and farsightedness). Notably, the feedback was only discovered in the AV case study, implying the acceptability of wearing AR glasses might depend on the envisioned frequency of use. Street crossing, in particular, is a daily activity, whereas travelling to a new location, requiring extensive use of navigational support, is a once-in-a-while occurrence. Qualitative results further revealed that participants---when assessing the wearable AR applications---differentiated between the perceived functional benefits that the application would provide and the technology involved. In other words, some participants found the functionalities beneficial, but at the same time, questioned whether AR glasses were the most appropriate technology for the job at hand. The application of AR glasses had to be sufficiently compelling to be used on a day-to-day basis and, to a significant extent, regarded as superior when compared to alternative technologies.

The post-study interviews showed that participants looked beyond experienced scenarios related to many real-world situations. For instance, participants in the AV study were interested not only in how the design concept aided their crossing but also in its effect on traffic efficiency. Interestingly, they voiced valid concerns about the concept's viability in mixed traffic conditions and data privacy, sharing similar viewpoints with scientific experts~\cite{tabone2021vulnerable}. These observations suggest that VR simulation could inspire a sense of an envisioned end-user experience, prompting participants to consider concepts for AR glasses more holistically, including aspects related to hardware, perceived functional benefits and the larger context. In other words, VR enabled the participants to immerse in the speculative future of AR glasses and their applications~\cite {simeone2022immersive}. This benefit is especially significant given that one of the challenges industry practitioners face in prototyping extended reality is bringing users into a world they have never experienced before~\cite{krauss2022elements}.

\subsubsection{Evaluating the Impact of Urban Contextual Factors}

Based on a review and clustering of the results relating to contextual factors, we determined the extent of their impact on user perception of AR applications.

\begin{itemize}
    \item(1) Safety concerns: The simulated context heightened the participants' awareness about their safety, prompting them to take note of the safety aspects of the design concepts (as in the navigation study) and their feeling of safety (as in the AV study). For instance, most participants in the navigation study expected to use the map interface \emph{while} walking and, therefore, disfavoured a design that might obscure the real world or distract them from their surroundings. As a result, researchers can utilise VR to experiment with different spatial arrangements of AR content or even give users the option to customise the placement in ways that best fit the situational context.
    
    \item(2) Attentional capacity: User statements suggested that the sheer quantity of visual distractions in the urban setting (e.g. two-way traffic) makes it more likely for them to overlook conformal AR contents, particularly those positioned at the periphery (e.g. car overlays). The divided attention exemplifies how VR prototyping may aid in the discovery of usability issues that might arise during in-situ deployments. Furthermore, simulated environmental distractions help determine whether the AR content might exceed the users' capabilities and cognitive demand in a specific situation. For example, the potential information overload in AV–pedestrian communication was examined in a multi-vehicle scenario, and the obtained feedback revealed the relevance of each information cue. Notably, much of the feedback was influenced by the temporal and spatial awareness of the virtual environment. For example, many participants expressed their fear of not knowing \emph{how close} the vehicles would stop from them, with one participant even hesitating to cross the road when the zebra crossing appeared \emph{before} the AVs had come to a complete stop. 

    \item(3) Social considerations: Participants were considerate of the potential adverse effects of their AR usage on others; for example, one participant in the AV study wanted to ensure that traffic movement would resume after he had crossed the street (A21). However, it should be noted that the number of social considerations related to interaction techniques was relatively small. This result could be attributed to the simple interaction in the AV study and the fact that only interactions with the on-hand map in the navigation study involved hand movement.
\end{itemize}

\subsection{VR Simulation Limitations (RQ2)}

Using the AR-within-VR method, designers and researchers can explore a number of design dimensions. However, as with other manifestations, it has some inherent limitations that should be considered. We believe it is essential to view VR simulation as a method with its own advantages and disadvantages and not as a replacement for high-fidelity AR prototypes and field research. In this section, we discuss the feedback that could not be anticipated when simulating wearable urban AR experiences in VR. 

\begin{itemize}

\item \textit{Examining visual fidelity:} Regarding AR visualisations, the prototypes conveyed a rather sophisticated appearance, inducing user feedback on numerous visual properties. However, the validity of the findings might be called into doubt when qualities such as colour, transparency, and text legibility are concerned. For example, while feedback about how the map was not transparent enough and the place-mark icons were hard to read aided in identifying usability issues, these observed issues may have been partly caused by the VR system (e.g. display resolution of the headset used). In addition, there may be issues that will not manifest under simulated conditions because the influencing factors, such as outdoor lighting~\cite{gabbard2006effects} and the visual complexity of background textures~\cite{merenda2019effects}, cannot be realistically recreated. Further, an interview study with industry practitioners reported that the targeted display technology could also tamper with the colours of an AR application~\cite{krauss2022elements}, necessitating the testing of colour variations via actual deployment on the target device. For these reasons, a simulated AR prototype may not be the most suited manifestation to examine the fine-grained visual qualities of AR interfaces and related usability issues (e.g. distinguishing colours \cite{hoggenmuller2021context}).

\item \textit{Producing exhaustive contextual feedback:} Although the VR simulation provided in-context feedback on wearable urban AR applications, it is important to note that the method should only be used as a first step to narrow down potential concepts and it does not replace field experiments for a number of reasons. First, prototyping and evaluating wearable AR applications in VR is a highly controlled activity by its very nature. This means that not only the application and the intended use environment are simulated, but also the instruction and tasks, resulting in a noisy approximation of how people actually use a product. In this regard, in-the-wild AR studies can be a more effective approach to understanding how an AR application is used in authentic everyday situations; for example, the participants in the study by~\citet{lu2021evaluating} used a working prototype in their own contexts for at least 2.5 hours unsupervised. Second, it is nearly impossible to replicate the messiness and variability of the real world and its potential effects. For example, in a field trial of a mobile AR map, \citet{morrison2009like} found that the sunlight and sunshade of the sun influenced the way participants held the device to view the AR visualisations. Therefore, VR bridges the gap between ecological validity and experimental control, but does not eliminate the `hassle' of conducting field studies \cite{kjeldskov2014worth, rogers2007s}.

\end{itemize}

\subsection{Recommendations for Simulating Wearable Urban AR Experiences in VR}

Building on the results of two case studies as the foundation and reflecting on our prototyping process, we distil and consolidate a set of preliminary recommendations (R) for simulating wearable urban AR experiences in VR. The recommendations focus on three dimensions that influence the efficacy of VR simulation, namely: (1) experience of wearing smartglasses, (2) AR content rendering, and (3) contextual setting. Examples from the two case studies are included throughout to illustrate the recommendations.

\subsubsection{Experience of Wearing Smartglasses}
When prototyping wearable AR applications, not only the mix of virtual and `real' content but also the experiential qualities of wearing spectacles are essential to consider. Although a VR headset does provide a similar experience to a certain extent, there is a need to reinforce the user's impression of wearing smartglasses. The purpose is not to assess the physical ergonomics of an AR system but to ensure that users (those unfamiliar with smartglasses) remain aware of the wearable technology employed and can conceive of its usage in everyday contexts. This emphasis applies especially when a study aims to compare non-AR interfaces with AR ones (as was the case in our AV study and a study by~\citet{prattico2021comparing}).

\begin{itemize}[label={},leftmargin=0pt]
\setlength\itemsep{0.5em}

\item 
\begin{mdframed}[hidealllines=true,backgroundcolor=gray!20]
\textbf{R1 - Emphasising the experience of wearing smartglasses}. Tapping is one of the most common hand gestures that allow users to engage with AR content. Smartglasses users can either perform air-tap gestures (as in HoloLens applications) or tap a touchpad on the temple (as in Google Glass applications). In the AV study, we implemented the latter mainly because it involves physical contact with the headset, which serves to emphasise the experience of wearing smartglasses. The gesture was natural, easy to implement, and contributed to greater user feedback on the AR glasses in the AV study. Yet, one disadvantage of this technique is that it is not applicable when AR interaction techniques~\cite{lee2018interaction, alce2015prototyping} are the focus of the investigation. 
\end{mdframed}
\end{itemize}

\subsubsection{AR Content Rendering.} Employing VR to simulate wearable AR resembles the Russian nesting doll effect, in which a UI is situated within another UI~\cite{alce2015prototyping}. Although we paid special attention to visual perceptions of AR content during the development process, the interaction effects remained strong and affected their recognisability, as evidenced by the qualitative feedback. It was sometimes difficult for participants to distinguish AR content from the virtual world, particularly when they were superimposed on `real' objects. Moreover, there was another issue with interaction effects between AR and VR that we did not anticipate: misunderstanding over system messages. Two participants were confused about whether the system feedback (an AR text prompt) was part of the actual AR experience or originated from the VR system. Given that wearable AR is an emerging technology that most people are still unfamiliar with, these issues are likely to occur and confound usability findings. To minimise the interaction effect between AR and VR content, AR augmentations should be rendered in a way that visibly separates them from the surrounding virtual environment that represents the `real' world.

\begin{itemize}[label={},leftmargin=0pt]
\setlength\itemsep{0.5em}
\item 
\begin{mdframed}[hidealllines=true,backgroundcolor=gray!20]

\textbf{R2 - Making AR content stand out:} To differentiate AR imagery from virtual environments, we created AR materials in emissive and translucent colours, resembling interfaces typically seen in science fiction films. To strengthen the effect, we propose increasing the realism of the environment with high-poly 3D models while using lower poly meshes for AR elements. However, because there is a trade-off between using complex mesh models and maintaining high simulation performance, particular attention must be paid to different VR optimisation techniques. 
\end{mdframed}

\item 
\begin{mdframed}[hidealllines=true,backgroundcolor=gray!20]
\textbf{R3 - Simulating registration inaccuracy:} A key issue of existing AR HMDs is the lack of image registration accuracy, which results in misalignments between the augmented imagery and physical objects~\cite{billinghurst2021grand}. While VR usage alleviates this problem, it was found during pilot tests that a perfect registration made distinguishing between AR and VR elements challenging. Therefore, we deliberately simulated a small degree of registration inaccuracy, for example, creating a noticeable gap between the car overlay and the car body. Participants in the pilot tests specifically commented on the usefulness of this technique in recognising digital overlays.
\end{mdframed}
\end{itemize}

\subsubsection{Contextual Setting.} Consideration should be given to the design of virtual environments to identify relevant circumstances that might influence user perceptions. While related literature and methods such as contextual observation are beneficial for this process, we found that pilot tests generated numerous insights for simulation iterations. For example, pilot test participants described a city without much traffic or people as \textit{`creepy'} and engaged in dangerous behaviours such as jaywalking. 

\begin{itemize}[label={},leftmargin=0pt]
\setlength\itemsep{0.5em}
\item 
\begin{mdframed}[hidealllines=true,backgroundcolor=gray!20]
\textbf{R4 - Determining the social and environmental factors to be incorporated:} Our findings demonstrate that the simulation of social and environmental factors frequently found in an urban setting, such as road traffic and human activities, contributed to participants' sense of presence. These factors are particularly critical when individuals are exposed to the virtual environment for an extended period (e.g. navigating or exploring the city). In addition to improving the experience in VR, however, the overall rationale for incorporating social and environmental factors should be to better assess their influence on the usability of urban AR applications. For example, in the navigation study, participants referring to background scenes we deliberately created (e.g. a policeman running after a thief) provided us with implicit but valuable feedback that our AR application offered sufficient situational awareness.  
\end{mdframed}

\item
\begin{mdframed}[hidealllines=true,backgroundcolor=gray!20]
\textbf{R5 - Incorporating different levels of details:} The extent to which contextual factors are modelled in detail, we argue, should vary according to their role and possible impact on the urban AR experience under investigation. Vehicle behaviour, for example, was not replicated as precisely in the navigation research as in the AV study because participants were not meant to engage directly with road traffic. Rather than managing every driving parameter (e.g. speed and deceleration rate), we used the built-in navigation system of Unity to fill the city with intelligent car agents, lowering the time and effort required to build the prototype. This also conforms with what \citet{lim2008anatomy} more broadly refers to as the economic principle of prototyping. 
\end{mdframed}

\end{itemize}


\subsection{Limitations and Future Work} 

According to \citet{dole2019face}, researchers can use measures such as users' sense of presence to assess a simulation's face validity, which is a proxy for ecological validity. In the selected case studies, participants' sense of presence was assessed through their statements and observed behaviours, rather than a standardised questionnaire, such as the Presence Questionnaire~\cite{witmer1998measuring}. To a large extent, the employed measures allowed us to ascertain specific elements that influence participants' sensations (e.g. soundscape and vehicles). However, questionnaires could potentially be useful in quantifying different dimensions of the construct. The immersion quality of the VR simulations was also constrained by the VR hardware used, the visual and interaction fidelity of the scenarios and the lack of self-presentation. While these issues may have impacted the sense of presence of the participants and, consequently, the quality of their feedback, they also present ample opportunities for VR simulations to become more effective in the future as VR technology advances~\cite{alce2015prototyping, simeone2022immersive} (e.g. a full body tracking would enable avatar legs).

The potential impact of social interactions was not apparent in the results, owing mainly to the case studies being designed to address their respective research questions. Social cues were incorporated into the virtual environment in ways that did not unnecessarily divert participants' attention away from the main experimental task. Future research should investigate whether the addition of virtual avatars in the same interaction space elicits increased input regarding social experience. Furthermore, similar to how \citet{flohr2020context} incorporated an actor as part of a shared ride video simulation, implementing social VR with real users may present an interesting research direction.

With respect to the validity of utilising VR to simulate AR, the literature has been concerned with two distinct aspects: the differences between the simulated and actual AR systems and those between the simulated and real-world environments. Regarding the former, several studies have reported initial evidence that the simulator results are equivalent to those obtained with real AR systems~\cite{lee2009replication, lee2010role, ragan2009simulation, bowman2012evaluating}. Regarding the latter, human behaviours and experiences in a virtual environment have been actively researched, with pedestrian simulators as one of the most prominent examples. According to a literature review conducted by~\citet{schneider2020virtually}, empirical data supporting generalisability to the real world exists but have been rather insufficient, and one should interpret such findings cautiously in terms of general trends rather than absolute validity. For these reasons, our study focused only on determining the relevance of obtained feedback to evaluating wearable AR experiences. 

AR hardware limitations and safety risks associated with our futuristic urban interfaces} have made it nearly impossible to compare simulated AR with real-life implementations. For instance, the current AR headsets do not have the wide FOV required for our navigation concepts, and it is difficult to mitigate the risk associated with interacting with AVs. Nevertheless, we believe it will be valuable to investigate the differences in user feedback by running follow-up field studies when the AR hardware and safety conditions are met. Furthermore, more simulation studies may and should be conducted to extend and refine the recommendations offered in this paper.


\section{Conclusion}
Wearable AR applications hold significant promise for transforming the relationship between individuals and urban environments. Furthermore, they have the potential to become a necessary component for interaction with emerging urban technologies, such as connected and cyber-physical systems in cities (e.g. AVs). An engaging wearable urban AR experience is closely linked to its perceived functional benefits and the context in which it is situated, both of which are essential to explore and assess throughout the design process. Through a comprehensive analysis of qualitative data from two wearable urban AR applications, this paper provides evidence demonstrating the potential of immersive VR simulation for evaluating a holistic and contextual user experience. The paper contributes to the body of knowledge in designing wearable urban AR applications in two specific ways. First, it offers empirically-based insights into the efficacy of VR simulation in terms of evaluating functional benefits and the impact of urban contextual factors. Second, the paper presents a set of recommendations for simulating wearable urban AR experiences in VR. We hope that our contributions will help overcome the barriers and complexities associated with prototyping and evaluating wearable urban AR applications. 

\begin{acks}
This research is supported by an Australian Government Research Training Program (RTP) Scholarship and through the ARC Discovery Project DP200102604, Trust and Safety in Autonomous Mobility Systems: A Human-centred Approach. 
\end{acks}

\bibliographystyle{ACM-Reference-Format}
\bibliography{ACM/references}

\end{document}